\documentclass[12pt]{article}
\pdfoutput=1 
\usepackage{graphicx,amsmath}
\usepackage{units}
\usepackage{color}
\usepackage{float}

\newlength{\dinwidth}
\newlength{\dinmargin}
\def\be{\begin{equation}}   
\def\ee{\end{equation}}  
\def\bea{\begin{eqnarray}}                      
\def\eea{\end{eqnarray}}
\def\ch1{$\chi(1^+)$}
\def\lapproxeq{\lower .7ex\hbox{$\;\stackrel{\textstyle                                                    
<}{\sim}\;$}}                                                    
\def\gapproxeq{\lower .7ex\hbox{$\;\stackrel{\textstyle                                                    
>}{\sim}\;$}}       
\def\cc{$c\bar{c}$}

\begin{document}

\begin{flushright}                                                    
IPPP/17/43 \\                                                    
\today \\                                                    
\end{flushright} 

\vspace*{0.5cm}

\begin{center}

{\Large \bf Low $x$ gluons determined by open charm production} \\
\vspace{0.5cm}
\vspace{1.0cm}

E.G. de Oliveira$^{a}$, A.D. Martin$^b$ and  M.G. Ryskin$^{b,c}$\\ 

\vspace{0.5cm}
$^a$ {\it Departamento de F\'{i}sica, CFM, Universidade Federal de Santa
Catarina, C.P. 476, CEP 88.040-900, Florian\'opolis, SC, Brazil}\\    
$^b$ {\it Institute for Particle Physics Phenomenology, University of Durham, Durham, DH1 3LE } \\
$^c$ {\it Petersburg Nuclear Physics Institute, NRC Kurchatov Institute, Gatchina, St.~Petersburg, 188300, Russia } \\ 

\begin{abstract}

We use the LHCb data on the forward production of $D$ mesons at 7, 13 and 5 TeV to make a {\it direct} determination of the gluon distribution, $xg$, at NLO in the $10^{-5}\lapproxeq  x \lapproxeq 10^{-4}$ region.
We first use a simple parametrization of the gluon density in each of the four transverse momentum intervals of the detected $D$ mesons. Second, we use a double log parametrization to make a combined fit to all the LHCb $D$ meson data. The values found for $xg$ in the above $x$ domain are of the similar magnitude (or a bit larger) than the central values of extrapolations of the gluons obtained by the global PDF analyses into this small $x$ region.  However, in contrast, we find $xg$ has a weak dependence on $x$.

\end{abstract}

\end{center}
\vspace{0.5cm}

\section{Introduction}

\begin{figure} [h]
\begin{center}
\includegraphics[clip=true,trim=0.0cm .0cm 0.0cm 0.0cm,width=12.0cm]{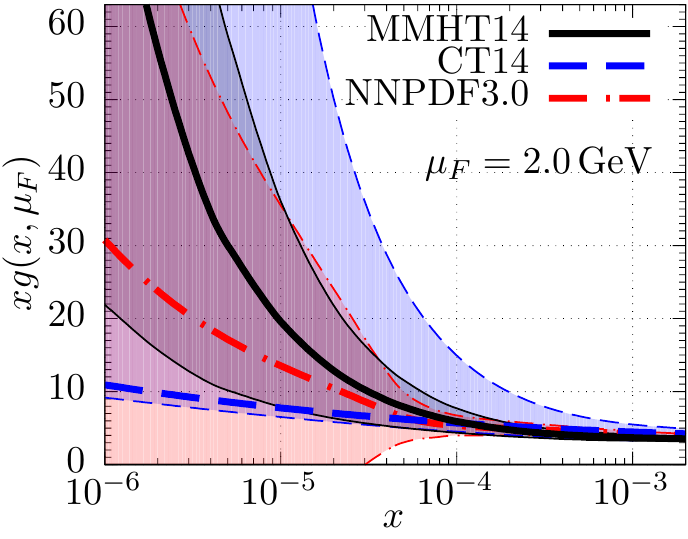}
\caption{\sf The low $x$ gluon density as given by recent NLO global parton analyses \ NNPDF3.0~\cite{Ball:2014uwa},
MMHT2014~\cite{Harland-Lang:2014zoa}, CT14~\cite{Dulat:2015mca}, calculated using the PDF interpolator LHAPDF~\cite{Buckley:2014ana}.}
\label{fig:g}
\end{center}
\end{figure}

The LHCb collaboration \cite{LHCbcc7,LHCbcc13,LHCbcc5} published the cross section for $D$ meson production in the forward direction with rapidities in the region $2<y<4.5$. These data can be described by the production of a 
$c\bar c$-pair followed by the fragmentation of  the $c$ quark into the $D$ meson. Moreover, gluon fusion, $gg$, is the major contributor to forward $D$ meson production. Therefore,  since the  mass of the $c$ quark, $m_c$, is not too high and that there is a large rapidity, this process allows a probe of the gluon distribution at very small $x$ \cite{R1,R2,R3,GR}
\be
x~\sim~ (m_T/\sqrt{s}) e^{-y}~ \sim ~10^{-5}
\ee
and at a small scale $\mu \sim m_T$. Here $m_T=\sqrt{m^2_c +p_{t,c}^2}$, where $p_{t,c}$ is the transverse momentum of the $c$ quark and $\sqrt{s}$ is the centre-of-mass proton-proton energy.  Due to the absence of data probing the gluon in the low $x$, low scale domain the gluon PDF is practically undetermined  in this domain by the global PDF analyses. The uncertainty at a scale $\mu=2$ GeV is illustrated in Fig.~\ref{fig:g}.

Formally, the inclusive cross section for $D$ meson production is calculated as a convolution of the hard matrix element with two parton PDFs and the $c\to D$ fragmentation function, which symbolically is of the form
\be
\frac{d^2\sigma}{dp_{t,D}~ dy}~=~{\rm PDF}(x_1,\mu_F) \otimes |{\cal M}|^2\otimes {\rm PDF}(x_2,\mu_F) \otimes D(z)
\label{eq:con}
\ee
where $z=p_{t,D}/p_{t,c}$ is the ratio of the $D$ and $c$ transverse momenta.  The momentum fractions $x_1,~x_2$ of the colliding partons are given by
\be
x_{1,2}~=~\frac{M_{c{\bar c}}}{\sqrt{s}}~e^{\pm Y}
\ee
where $M_{c{\bar c}} $ and $Y$ are the mass and the rapidity of the \cc ~system: that is
\be
M_{c{\bar c}}~=~\sqrt{p_{t,c}^2 +m_c^2~ }~ 2~{\rm cosh}(\Delta Y/2),~~~~~~~Y=(y_c+y_{{\bar c}})/2~,
\ee
and $\Delta Y=(y_c-y_{{\bar c}})~$.

A problem is that it is not completely clear what factorization scale, $\mu_F$, should be used in (\ref{eq:con}). On the other hand, in the low $x$ region the gluon density strongly depends on the scale $\mu_F$ due to the presence of  
Double Log (DL) terms $(\alpha_s N_C/\pi){\rm ln}(1/x){\rm ln}\mu_F^2$ with a large $\ln(1/x)\sim 10$. This may be one of the reasons\footnote{Compared to other data used in the global PDF analyses \cite{Ball:2014uwa,
Harland-Lang:2014zoa,Dulat:2015mca}, the number of events measured for forward open charm production is relatively very small and consequently, at present, the measured $c\bar{c}$ cross sections have much lower statistical weight.} why such data are not yet used in global analyses. It was shown in \cite{OMR} that the DL terms can be resummed in the incoming parton PDFs by choosing $\mu_F \simeq 0.85m_T$. After fixing the value of $\mu_F$ in the LO part of the cross section, the remaining dependence on the scale $\mu_f$ becomes weaker. Recall that at NLO the cross section can be written as 
\begin{equation}
\sigma^{(0)}(\mu_f) + \sigma^{(1)}(\mu_f) ~=~\alpha_s^2\left[{\rm PDF}(\mu_F)\otimes C^{(0)} \otimes {\rm PDF}(\mu_F) +
{\rm PDF}(\mu_f)\otimes\alpha_s C_{\rm rem}^{(1)}(\mu_F) \otimes {\rm PDF}(\mu_f)\right]~,  
\label{eq:stab1}
\end{equation}
where all the $\mu_f$ dependence of cross section on the left-hand-side comes from the NLO term on the right-hand-side. The contribution of the remaining NLO term is much less than the LO part where the scale is fixed. Now, with rather good accuracy, we can say that open charm production probes PDFs at a known scale $\mu_F\simeq 0.85m_T$.

\section{Description of the data}
The major contribution to open \cc~production comes from the  $gg$ fusion subprocess in (\ref{eq:con}). 
Here we fit the LHCb measurements\footnote{Note that the measurements at 13 and 5 TeV have recently been corrected. We use these updated data.} of this process obtained at 7, 13, 5 TeV \cite{LHCbcc7,LHCbcc13,LHCbcc5} to determine the low $x$ gluon distribution. The data, $d^2\sigma/dp_{t,D}dy$, are measured for five $D$ meson rapidity intervals in the range $2<y<4.5$ and we use four transverse momentum bins covering the range  $1<p_{t,D}<5$ GeV. We fit to the data for $D^\pm, ~D^0, ~\bar{D}^0$ meson production.  For these LHCb data the typical value of $x_1$ \gapproxeq 0.01 for one gluon is much larger than the value $x_2 \sim 10^{-5}$ of the other. Since for $x\gapproxeq 0.01$ the uncertainties of the global analyses are small we can take the larger $x$ PDFs from the global analyses, while we fit the low $x$ gluon PDF to the data. In addition to pure $gg$  fusion, there will be $gq$ and $q{\bar q}$ contributions. For these smaller terms we take the quark PDFs from the global analyses.    The $c\to D$ fragmentation functions $D(z)$ was taken from \cite{cac}, where they were determined from $e^+e^-$ annihilation data in the $\Upsilon(4S)$ region.

Actually the relative normalization of $c$ quark fragmentation to the different channels is only known from previous data to about 10$\%$ accuracy.  Therefore we allow, via a parameter $N^D$, for an additional renormalization of the $D^0,{\bar D}^0$ relative to the $D^\pm$ data.

The FONLL programme~\cite{FONLL} was used to calculate the open charm cross section at NLO.

\section{Determination of the low $x$ gluon}
Here we fit to the LHCb data for $D^\pm, ~D^0, ~\bar{D}^0$ open charm production \cite{LHCbcc7,LHCbcc13,LHCbcc5} in order to determine the low $x$ gluon distribution. As mentioned above, we work at NLO. The running coupling, the charm mass (1.4 GeV), the quark and high $x$ gluon distributions are given by the results found by the MMHT2014 NLO global fit.

\subsection{Gluons at fixed scales with a simple parametrization  \label{sec:3.1}}
The data in each of the four $p_t$ intervals ($1-2,~2-3,~3-4,~4-5$ GeV) were fitted separately assuming a simple two-parameter form for the low  $x$ behaviour of the gluon
\be
xg(x)=N\left(\frac x{x_0}\right)^{-\lambda}
\label{eq:Nlam}
\ee
with $x=x_2$ and $x_0=10^{-5}$.  In this way we obtain gluons at four values of $\mu_F$.  Note that the value of $\mu_F$ is a bit larger than $\sqrt{p_{t,D}^2+m_D^2}$ since after fragmentation the transverse momentum of the $c$ quark 
\be
p_{t,c} ~=~p_{t,D}/z~ >~ p_{t,D}.
\ee
In fact $\mu_F= 2.0, ~2.9, ~3.9, ~4.9$ GeV for $p_{t,D}=1.5,~2.5,~3.5,~4.5$ GeV respectively. 

As mentioned above, we allowed an extra normalization parameter, $N^D$, between the $D^0,{\bar D}^0$ and the $D^\pm$ data. It was found to be
 $N^D\simeq 1.1$ in every case. In detail, this means we take the fractions 0.246 and 1.1(0565) for the $D^\pm$ and $(D^0+{\bar D}^0)$ charm quark decay channels respectively, leaving 0.133 for the $D_s+\Lambda_c$ channels. The results are essentially unchanged if we set $N^D=1$, but then $\chi^2$ is a bit larger.
 
\begin{figure} [!htb]
\begin{center}
\vspace{-0.0cm}
\includegraphics[clip=true,trim=0.0cm .0cm 0.0cm 0.0cm,width=8.2cm]{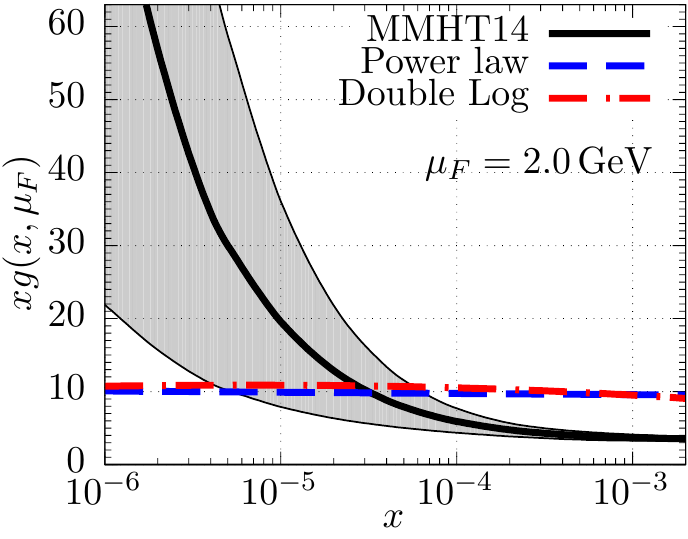} \hspace{.2cm}
\includegraphics[clip=true,trim=0.0cm .0cm 0.0cm 0.0cm,width=8.2cm]{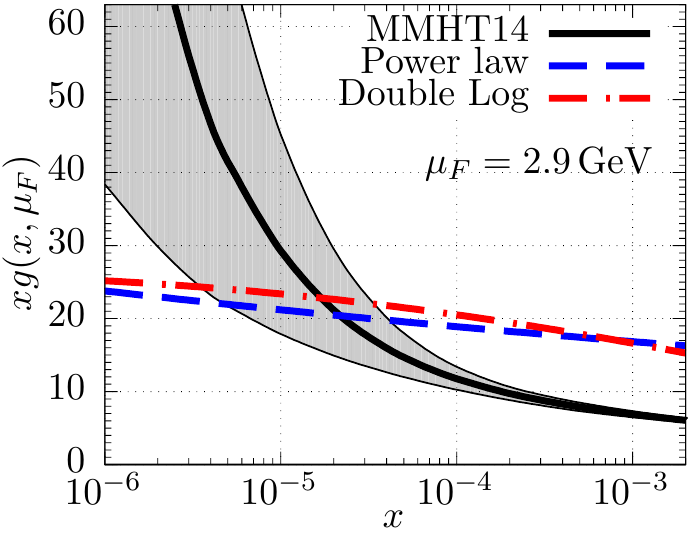}
\includegraphics[clip=true,trim=0.0cm .0cm 0.0cm 0.0cm,width=8.2cm]{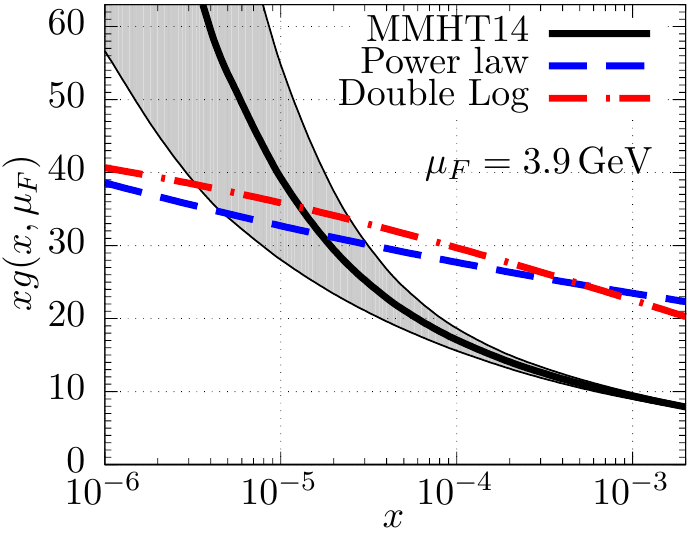} \hspace{.2cm}
\includegraphics[clip=true,trim=0.0cm .0cm 0.0cm 0.0cm,width=8.2cm]{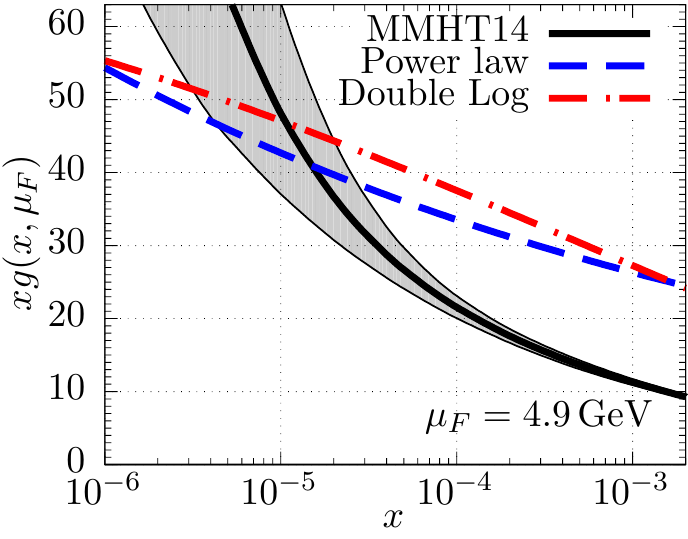}
\caption{\sf The four plots show the low $x$ behaviour of the gluon distribution in the four different $p_t$ intervals, obtained by two different fits to the LHCb open charm data \cite{LHCbcc7,LHCbcc13,LHCbcc5}. To be precise, the dashed curves are the gluon PDFs obtained by fitting to the data in each $p_{t,D}$ interval individually,  whereas the dash-dotted curves are the result of the `combined' fit. These two sets of curves are compared with the gluon PDFs extrapolated using the MMHT14 NLO global parton analysis \cite{Harland-Lang:2014zoa}, with uncertainties shown by the shaded bands.}
\label{fig:plots}
\end{center}
\end{figure}

 The normalization $N$ and the power $\lambda$ in (\ref{eq:Nlam}) and their uncertainties are obtained by fitting to the data in each $p_{t,D}$ interval using MINUIT numerical minimization code~\cite{minuit,James:2004xla}.
The results of the four fits are presented in Table \ref{tab:1} and Fig. \ref{fig:plots}.
\begin{table}[hbt]
\label{tab:diff}
\begin{center}
\begin{tabular}{|c|c|c|c|c||c|c|c|}\hline
$p_{t,D}$ & $\mu_F$ & $N$ & $\lambda $ & $\chi^2_{\rm all} $ &  $\chi^2_5$ & $\chi^2_7 $ & $\chi^2_{13}$  \\

\hline
1.5&2.0& $9.9\pm0.4$&$0.01\pm 0.01$&51& 19 & 6 & 26 \\
  2.5& 2.9 & $21.2\pm 0.8$&$0.05\pm 0.01$&31& 11 & 6 & 13 \\
  3.5 & 3.9 & $32.7\pm 1.5$&$0.07\pm 0.01$&27& 7 & 12 & 8 \\
4.5 & 4.9& $42.7\pm 1.5$&$0.10\pm 0.01$&29& 4 & 14 & 11 \\
 \hline
\end{tabular}
\caption{\sf The parameters $N$ and $\lambda$ giving the low $x$ behaviour of the gluon distribution, $xg(x,\mu_F)=N(x/x_0)^{-\lambda}$ from individual fits to the LHCb open charm data \cite{LHCbcc7,LHCbcc13,LHCbcc5} in the four different $p_{t,D}$ intervals.  The scale $\mu_F$ and $p_{t,D}$ are given in GeV. The total $\chi^2$, $\chi^2_{\rm all}$, in each interval is shown, together with the contributions from the 5, 7 and 13 TeV data.}
\label{tab:1}
\end{center}
\end{table}

From the individual $\chi^2$ contributions in Table \ref{tab:1} we see hints of tension between the 5+13 TeV data on the one hand and the 7 TeV on the other hand. Surprisingly if the 5+13 TeV data are fitted without the 7 TeV data (and vice-versa) we find similar values of $N$ and $\lambda$, although the values of $\chi^2$ are reduced.

\subsection{Combined fit using a Double Log (DL) parametrization}
As seen from the results of the simple fits shown in Table \ref{tab:1} the gluon density increases strongly with scale, while the power of the $x$ behaviour has a weaker scale dependence. It is not evident whether such a behaviour is consistent with DGLAP evolution. Since in the low $x$ region the major effect comes from the DL contribution, we fit all four groups of data simultaneously with the formula
\be
xg(x,\mu^2)~=~N^{\rm DL}\left(\frac x{x_0}\right)^{-a}\left(\frac{\mu^2}{Q_0^2}\right)^b{\rm exp}\left[\sqrt{16(N_c/\beta_0){\rm ln}(1/x){\rm ln}(G)}\right]
\ee
with
\be
G=\frac{{\rm ln}(\mu^2/\Lambda^2_{\rm QCD})}{{\rm ln}(Q_0^2/\Lambda^2_{\rm QCD})}.
\label{eq:gPDF}
\ee
With three light quarks $(N_f=3)$ and $N_c=3$ we have $\beta_0=9$. The resummation of the leading double logarithmic terms $(\alpha_s{\rm ln}(1/x){\rm ln}(\mu^2))^n$ is written explicitly in the exponential, while the remaining single log terms are now accounted for by the powers $a,~b$. We take $\Lambda_{\rm QCD}=200$ MeV and $Q_0 =1$ GeV. As before we take $x_0=10^{-5}$.   

Such an ansatz was used successfully to describe $J/\psi$ and $\Upsilon$ photoproduction data \cite{JMRT}. Moreover, it was checked that in the $x,~\mu_F^2$ region of interest this formula was consistent with NLO DGLAP evolution. So we fix the power $b$, which is responsible for the $\mu_F^2$ behaviour, to be the same as that found in the fit to $J/\psi$ photoproduction. Now  we are left to describe 120 LHCb data point with only two free parameters:  $N^{\rm DL}$ and $a$, plus the parameter $N^D$ introduced in Section~\ref{sec:3.1}.

\begin{table}[hbt]
\label{tab:diff}
\begin{center}
\begin{tabular}{|c|c|c|c|c||c|c|c|}\hline
&  $N^{\rm DL}$ & $a$ &$b$ (fixed)& $\chi^2$  &  $\chi^2_5$ & $\chi^2_7 $ & $\chi^2_{13}$  \\
\hline
Fit to \cc~data&$0.13\pm 0.01$ & $-0.20\pm 0.01$ &$-0.2$&141 & 44 & 40 & 56\\
Fit\cite{JMRT} to $J/\psi$ data& $0.092\pm 0.009$& $-0.10\pm 0.01$&$-0.2$ & & & &\\
  \hline
\end{tabular}
\caption{\sf The values of the parameters $N^{\rm DL}$ and $a$ obtained in a fit to all the LHCb open charm data \cite{LHCbcc7,LHCbcc13,LHCbcc5} (120 data points in total) using the DL parametrization given by eq.(\ref{eq:gPDF}). We also show what the three data sets contribute to the total $\chi^2=141$.  For comparison we show the parameters of the gluon obtained in a similar fit \cite{JMRT} to $J/\psi$ data. Note, however, that the gluons obtained in~\cite{JMRT} from $J/\psi$ data are not the $\overline{\mbox{MS}}$ gluons but corresponds to another (physical) factorization scheme.}
\label{tab:2}
\end{center}
\end{table}




The results of the combined fit are presented in Table \ref{tab:2},
 and are compared in Fig.~\ref{fig:plots} with the results of the simple fits. 
 We see that we have an acceptable description of all the $D$ meson data.
From Fig.~\ref{fig:plots} we see that the gluons from the simple fit and the DL fit coincide for $x\simeq 2\times 10^{-5}$, which shows the region where the body of the open charm data probes the gluon PDF. The values of the gluon density found in this domain, $10^{-5}\lapproxeq x \lapproxeq 10^{-4}$, represent the first direct determination from data. The DL description, which embodies NLO DGLAP evolution, should be more reliable than the results of the simple fits.

Note that as we enter the $x\lapproxeq 10^{-4}$ domain we have a direct determination of $xg$ which has a flatter $x$ behaviour than that of the gluon density of the global PDF analyses extrapolated down to this domain.

There is some tension between the rapidity behaviour of D-meson cross sections at a fixed energy and the dependence of cross sections on initial proton-proton energy. In particular, one may consider the ratio of D-meson yields taking at rapidities shifted by $\delta y=y_1-y_2=\ln(s_1/s_2)$ where $s_1$ and $s_2$ are the energy squared in two different runs. In such a case the value of large $x_1$ will be exactly the same and the ratio of cross sections will be equal just to the ratio of gluon densities taken at fixed scale, $\mu_F$, and different $x_2$. Surprisingly in all 4 $p_t=1.5, 2.5, 3.5, 4.5$ GeV bins within the errors these ratios are well described by the constant $xg(x,\mu_F)=\text{const}(x)$, that is by $\lambda=0$ in parametrization (\ref{eq:Nlam}).

\section{Conclusions}

Inclusive $D$ meson cross section data obtained in the forward region by LHCb  allow\cite{LHCbcc7,LHCbcc13,LHCbcc5}, for the first time, a direct probe of the gluon density at low $x$ and low scales. Including these data in the global PDF analyses will crucially decrease the uncertainty of the present PDFs in this low $x$ domain.  In this paper, we did not perform a global PDF analysis but concentrated on establishing the gluon PDF needed to describe the open \cc~production measurements. This way we obtain the low $x$ gluons directly measured in this region by LHCb experiment not affected by the constraints coming from a particular form of ansatz used to describe the input PDFs.

How do our results compare with previous determinations of the low $x$ gluon density from open charm production data? The latest such determination is given in \cite{GR}.  Apart from the fact that we have been able to use
 the most recent {\bf corrected} data on the $D$ meson cross sections, the most important difference is that in~\cite{GR} only the {\em ratios} of inclusive cross sections were fitted,  whereas we fit to the {\it absolute} cross section data.  Of course the ratios are more stable and less dependent on the choice of the scale. On the other hand, using only ratios one cannot determine the normalisation of the gluon PDFs. Actually in such a case the normalization is given by the matching with the larger $x$ PDFs obtained by the global PDF analyses. Therefore it is not so surprising that we get more or less the same (rather flat) $x$-behaviour (relatively small $\lambda\sim 0.01$ at 
$\mu_F=2.0$ GeV)
 but about twice as large a normalization.

An earlier analysis~\cite{R2} incorporated the 7 TeV $c{\bar c}$ (and $b{\bar b}$) data in a `global' fit including HERA structure function data. Two `global' fits were performed using either absolute or normalized heavy quark data. The `normalized' fit suffers from the same defect as that in~\cite{GR}. In the `absolute' fit the factorization and renormalization scales were allowed to be free parameters. The values found for the scales are somewhat lower than the optimal choice $\mu_F(=\mu_R)=0.85 m_T$. In particular they found $\mu_R=0.44m_T$. This low renormalization scale was needed to enlarge the charm quark cross section. After this the magnitude of the gluon in the range $10^{-5}<x<10^{-4}$ at $\mu^2_F=10$ GeV$^2$ found in \cite{R2} are comparable to those in Fig.~\ref{fig:plots}.

The cross section data for forward open charm production have, at present, relatively low statistics and consequently much lower statistical weight in comparison with the other data used in global parton analyses \cite{Ball:2014uwa, Harland-Lang:2014zoa, Dulat:2015mca}. On the other hand, they directly probe the gluon PDF in the range, $10^{-5}\lapproxeq  x \lapproxeq 10^{-4}$, not covered by other data.
In this unexplored domain, we find that the gluon densities are of similar magnitude (or a bit larger) 
as compared to the central values obtained by extrapolations of the gluon densities of the global  analyses,  However, in contrast to the extrapolated global gluon PDFs, we find that the direct measurements of $xg$ have a weaker dependence on $x$ in this domain.  Higher statistics forward charm data will be valuable to resolve this dilemma.

\section*{Acknowledgement}
We are grateful to Ronan McNulty for valuable discussions. 
MGR thanks the IPPP at Durham University for hospitality and
  the RSCF grant 14-22-00281 for support. EGdO was supported by Capes and CNPq (Brazil).


\thebibliography{}

\bibitem{LHCbcc7} R. Aaij {\it et al.} [LHCb Collaboration], Nucl. Phys. {\bf B871} (2013) 1.   
\bibitem{LHCbcc13}  R.~Aaij {\it et al.} [LHCb Collaboration], JHEP {\bf 1603} (2016), 159. Erratum: JHEP {\bf 1609} (2016), 013; JHEP {\bf 1705} (2017), 074.

\bibitem{LHCbcc5} R.~Aaij {\it et al.} [LHCb Collaboration], arXiv:1610.02230v5.
 
\bibitem{R1} R. Gauld, J. Rojo, L. Rottoli and J. Talbert, JHEP {\bf 11} (2015) 009. [arXiv:1506.08025].

\bibitem{R2} O. Zenaiev {\it et al.}, Eur. Phys. J. {\bf C75} (2015) 396 [arXiv:1503.04581].

\bibitem{R3} M. Cacciari, M.L. Mangano and P. Nason, Eur. Phys. J. {\bf C75} (2015) 610 [arXiv:1507.06197].
 
\bibitem{GR}  R. Gauld and J. Rojo, Phys. Rev. Lett. {\bf 118} (2017) 072001  [arXiv:1610.09373].
 
%
\bibitem{Ball:2014uwa}
  R.D.\ Ball {\it et al.}  [NNPDF Collaboration],
  JHEP {\bf 1504} (2015) 040.
%
\bibitem{Harland-Lang:2014zoa}
  L.A.\ Harland-Lang, A.D.\ Martin, P.\ Motylinski and R.S.\ Thorne,
  Eur.\ Phys.\ J.\ {\bf C75} (2015) 5, 204.
%
\bibitem{Dulat:2015mca}
  S.\ Dulat, T.J.\ Hou, J.\ Gao, M.\ Guzzi, J.\ Huston, P.\ Nadolsky,
  J.\ Pumplin, C.\ Schmidt, D. Stump and C.P. Yuan, Phys. Rev. {\bf D93} (2016) 033006 [{\tt arXiv:1506.07443}].

\bibitem{Buckley:2014ana} 
A.~Buckley, J.~Ferrando, S.~Lloyd, K.~Nordström, B.~Page, M.~Rüfenacht, M.~Schönherr and G.~Watt,
Eur.\ Phys.\ J.\ C {\bf 75}, 132 (2015)
doi:10.1140/epjc/s10052-015-3318-8
[arXiv:1412.7420 [hep-ph]].

%
 \bibitem{OMR}E.G. de Oliveira, A.D. Martin and M.G. Ryskin, Eur. Phys. J. {\bf C77} (2017) 182 [{\tt arXiv:1610.06034}].
 
 \bibitem{cac} M. Cacciari, P. Nason and C. Oleari, JHEP {\bf 0604} (2006) 006 [hep-ph/0510032]. 

\bibitem{FONLL} S. Forte, E. Laenen, P. Nason and J. Rojo, Nucl. Phys. {\bf 834} (2010) 116.

\bibitem{minuit} F. James and M. Roos, Compt. Phys. Commun. {\bf 10}, (1975) 343.

\bibitem{James:2004xla}
F.~James and M.~Winkler,
``MINUIT User's Guide.''


\bibitem{JMRT} S.P. Jones, A.D. Martin, M.G. Ryskin and T. Teubner, J. Phys. G {\bf 44} (2017) 03LT01 [arXiv:1611.03711].
\end{document}